# The Extended Resonant Modal Theory and Its Applications

Ruqi Xiao, Wen Geyi, *Fellow, IEEE*, Guo Yang, and Wen Wu, *Senior Member, IEEE*

*Abstract*—In this paper, we extend the resonant modal theory (RMT) developed previously for a metal object to an arbitrary source region consisting of metals, dielectrics, or the combination of both. The influences of dielectrics on the fields are replaced by equivalent volume sources through the use of the compensation theorem in electromagnetic theory. The resonant frequencies can be determined by finding the roots of the determinant of the matrix resulted from the discretization of the real homogeneous volume-surface integral equation derived from the requirement that the difference of stored field energies in the source region vanishes. As applications of the extended RMT, three examples have been investigated. The first example is a dielectric resonator antenna, and is designed by exciting the first resonant mode of the composite structure in which the dielectric cylinder is combined with a conformal metallic strip. The second example is a dual-band dielectric-coated metallic wire antenna. The third example studies the resonant modes of a rectangular patch antenna.

## I. INTRODUCTION

Physically, an arbitrary scatterer (or source region) is said to be resonant if the stored electric and magnetic field energies of the scatterer are equal [1]-[3]. The external resonant modes for the scatterer, corresponding to the real resonant frequencies, are commonly used in antenna designs and scattering analysis. There have been several modal theories used to investigate the external resonant modes, including the singularity expansion method (SEM) [4], the eigenmode expansion method (EEM) [5], and the theory of characteristic mode (TCM) [6]-[9]. The three modal analysis methods result in three different kinds of modes, the natural resonant modes, the eigenmodes, and the characteristic modes, respectively. Moreover, the resonant modal theory (RMT) has been proposed in [10] to evaluate the external resonant modes of a metal scatterer.

The formulations of SEM, EEM, and TCM are all rely on the relationship between the scattering and incident fields. For the

This work was supported by the Chinese Natural Science Foundation (Grant No. 61971231). (Corresponding author: Wen Geyi)

Ruqi Xiao, Guo Yang and Wen Wu are with the School of Electronic and Optical Engineering, Nanjing University of Science and Technology, Nanjing 210094, China (e-mail: Xiaoruqi_ee@njust.edu.cn).

Wen Geyi is with the Research Center of Applied Electromagnetics, Nanjing University of Information Science and Technology, Nanjing 210044, China (e-mail: wgy@nuist.edu.cn).

SEM in [11]-[12] and the impedance-based TCM in [8], [13]-[16], different boundary conditions are utilized in combination with the relationship between the scattering and incident fields to construct various integral equations, which are then employed to generate different impedance matrices using the method of moments (MoM). The scattering-based TCM aims to directly calculate the scattering dyadic (transition matrix) between the scattering and incident field through various numerical methods [17]-[18], but treatments of boundary conditions are also necessary and important for the numerical accuracy of the scattering dyadic, as emphasized in [18]. Different from the above-mentioned modal theories, a homogeneous surface integral equation (SIE) is derived to determine the resonant modes of metals in [10] by requiring that the difference of stored field energies (DSFE) vanishes, which avoids the treatments of boundary conditions.

The natural resonant modes of SEM [4], [11]-[12] and the eigen-modes of EEM [5] are all determined in the complex frequency domain, which consumes considerable computational time. The modal analysis for SEM is carried out by searching the roots of the determinant of impedance matrices, and the root seeking along the imaginary frequency axis introduces heavy computation burden [19]. Different from SEM and EEM, the TCM is carried out in the real frequency domain [6]-[9], of which the characteristic values and characteristic modes are all real and calculated from various eigenvalue equations. After that, one needs to sort and track the characteristic values in a wide frequency range based on the hypotheses of various metrics of modal similarity, known as modal tracking, and only the frequencies with the characteristic values close to zero can be considered as the resonant frequencies [20]. The modal tracking could be a challenge sometimes [20]-[21]. Similar to the TCM, the RMT in [10] is also conducted in real frequency domain. However, the RMT is to determine the resonant modes of metals through finding the roots of the determinant of the real homogeneous algebraic equation using bisection method in real frequency axis, which avoids complex numerical implementations of the root seeking along the imaginary frequency axis required for SEM and the modal tracking necessary for TCM.

In this paper, the RMT is extended to an arbitrary source region consisting of conductors, dielectrics, or their combinations. Compared with the RMT developed in [10], which only involves a real homogeneous SIE for determining the resonant modes of conductors, the extended RMT in this work has considered the influences of dielectrics on the fields, which are replaced by an equivalent volume source through the compensation theorem. The new process yields a real homogeneous volume-surface integral equation (VSIE) that can be employed to determine the resonant modes of metal, dielectric, or the combination of metal and dielectric object. Distinguished from the other modal theories, the RMT are derived by requiring that DSFE in the source region must vanish, and the resonant modes for arbitrary source region can be determined easily by finding the roots of a determinant in the real frequency domain.

The rest of the paper is organized as follows. Section II discusses the formulation of the extended RMT and the numerical procedures for determining the resonant modes in an arbitrary source region. Section III shows three different applications of the RMT in antenna designs. A summary is provided in Section IV.

## II. Extended RMT

The extended RMT consists of three steps: 1) The DSFE of an arbitrary source region is obtained from the Poynting theorem and compensation theorem of electromagnetic fields. 2) By requiring that the DSFE must vanish, a real homogeneous integral equation is obtained and is then discretized into a homogeneous algebraic equation. 3) The resonant modes are determined by the existence of non-trivial solutions for the homogeneous algebraic equation.

### A. Formulations of extended RMT

Fig. 1 shows an arbitrary source region $V_0$ bounded by $\partial V_0$, which contains a current distribution $\boldsymbol{J}$ in free space. Let $S$ be the boundary of a volume $V$ containing the source region $V_0$. Applying the complex Poynting theorem to the region $V$ yields [22]

$$-\frac{1}{2}\int_{V_0}\boldsymbol{J}^*\cdot\boldsymbol{E}\,\mathrm{d}V = \oiint_S\mathbf{u}_n\cdot\boldsymbol{S}\,\mathrm{d}S + \mathrm{j}2\omega\int_V\frac{1}{4}\left(\mu_0\left|\boldsymbol{H}\right|^2-\varepsilon_0\left|\boldsymbol{E}\right|^2\right)\mathrm{d}V, \tag{1}$$

where $\mathrm{j}=\sqrt{-1}$; $\omega$ is the angular frequency; $\boldsymbol{S}=\boldsymbol{E}\times\boldsymbol{H}^*/2$ is the complex Poynting vector; $\boldsymbol{H}$ and $\boldsymbol{E}$ represent the magnetic and electric fields produced by the source current distribution $\boldsymbol{J}$; $\mu_0$ and $\varepsilon_0$ are the permeability and permittivity in free space; and superscript $*$ designates the complex conjugate operation. Note that the RMT formulations in [10] can only be applied to determine the resonant modes of metals. In this section, the RMT will be extended to an arbitrary source region consisting of metals, dielectrics, or the combination of both.

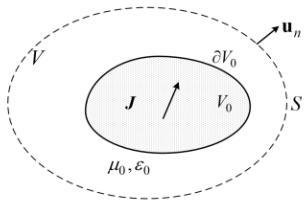

Fig. 1 An arbitrary source region.

Based on the compensation theorem of electromagnetic field [3], [23], the influence of any dielectric with permeability $\mu_0$

and permittivity $\varepsilon_\mathrm{d}$ on the field can be replaced by an equivalent volume source

$$\boldsymbol{J}_\mathrm{V} = \mathrm{j}\omega\left(\varepsilon_\mathrm{d}-\varepsilon_0\right)\boldsymbol{E}. \tag{2}$$

The current $\boldsymbol{J}$ in $V_0$ is thus divided into the current $\boldsymbol{J}_\mathrm{s}$ on the conducting surface and the equivalent volume current $\boldsymbol{J}_\mathrm{V}$ in the dielectric region. Due to the introduction of the dielectric (variation of medium parameters) into the source region, the DSFE in $V$ thus can be written as

$$\Delta W = \int_{V-V_0}\frac{1}{4}\left(\mu_0\left|\boldsymbol{H}\right|^2-\varepsilon_0\left|\boldsymbol{E}\right|^2\right)\mathrm{d}V$$
$$+ \int_{V_0}\frac{1}{4}\left(\mu_0\left|\boldsymbol{H}\right|^2-\varepsilon_\mathrm{d}\left|\boldsymbol{E}\right|^2\right)\mathrm{d}V \tag{3}$$
$$= \int_V\frac{1}{4}\left(\mu_0\left|\boldsymbol{H}\right|^2-\varepsilon_0\left|\boldsymbol{E}\right|^2\right)\mathrm{d}V - \int_{V_0}\frac{1}{4}\left(\varepsilon_\mathrm{d}-\varepsilon_0\right)\left|\boldsymbol{E}\right|^2\,\mathrm{d}V.$$

In (3), the DSFE is divided into two terms after the second equal sign: the first term is the energy difference when the region $V$ is free space, and the second term represents the energy difference caused by the variation of medium parameters in the source region. Note that the equivalent volume source in (2) and the second term after the second equal sign of (3) vanish if there are no dielectric introduced in the source region. The first term after the second equal sign of (3) can be obtained through (1) as follows

$$\int_V\frac{1}{4}\left(\mu_0\left|\boldsymbol{H}\right|^2-\varepsilon_0\left|\boldsymbol{E}\right|^2\right)\mathrm{d}V = \frac{1}{2\omega}\mathrm{Im}\left[-\frac{1}{2}\int_{V_0}\boldsymbol{J}^*\cdot\boldsymbol{E}\,\mathrm{d}V\right], \tag{4}$$

where Im represents the imaginary part of the complex vector. Same as in [10], the electric filed $\boldsymbol{E}$ in (4) can be represented by the scalar and vector potential functions using the surface current $\boldsymbol{J}_\mathrm{s}$ and equivalent volume current $\boldsymbol{J}_\mathrm{V}$. Therefore, the first term after the second equal sign of (3) can be rewritten as

$$\int_V\frac{1}{4}\left(\mu_0\left|\boldsymbol{H}\right|^2-\varepsilon_0\left|\boldsymbol{E}\right|^2\right)\mathrm{d}V$$
$$= \frac{\eta}{16\pi c}\int_{S_\mathrm{m}}\int_{S_\mathrm{m}}\left(\boldsymbol{J}_\mathrm{s}^*\cdot\boldsymbol{J}_\mathrm{s}-\frac{1}{k^2}\nabla\cdot\boldsymbol{J}_\mathrm{s}^*\nabla'\cdot\boldsymbol{J}_\mathrm{s}\right)\frac{\cos kR}{R}\,\mathrm{d}S'\mathrm{d}S$$
$$+ \frac{\eta}{16\pi c}\int_{S_\mathrm{m}}\int_{V_\mathrm{d}}\left(\boldsymbol{J}_\mathrm{s}^*\cdot\boldsymbol{J}_\mathrm{V}-\frac{1}{k^2}\nabla\cdot\boldsymbol{J}_\mathrm{s}^*\nabla'\cdot\boldsymbol{J}_\mathrm{V}\right)\frac{\cos kR}{R}\,\mathrm{d}V'\mathrm{d}S \tag{5}$$
$$+ \frac{\eta}{16\pi c}\int_{V_\mathrm{d}}\int_{S_\mathrm{m}}\left(\boldsymbol{J}_\mathrm{V}^*\cdot\boldsymbol{J}_\mathrm{s}-\frac{1}{k^2}\nabla\cdot\boldsymbol{J}_\mathrm{V}^*\nabla'\cdot\boldsymbol{J}_\mathrm{s}\right)\frac{\cos kR}{R}\,\mathrm{d}S'\mathrm{d}V$$
$$+ \frac{\eta}{16\pi c}\int_{V_\mathrm{d}}\int_{V_\mathrm{d}}\left(\boldsymbol{J}_\mathrm{V}^*\cdot\boldsymbol{J}_\mathrm{V}-\frac{1}{k^2}\nabla\cdot\boldsymbol{J}_\mathrm{V}^*\nabla'\cdot\boldsymbol{J}_\mathrm{V}\right)\frac{\cos kR}{R}\,\mathrm{d}V'\mathrm{d}V,$$

where $R=\left|\boldsymbol{r}-\boldsymbol{r}'\right|$, $\eta=\sqrt{\mu_0/\varepsilon_0}$, $c=1/\sqrt{\mu_0\varepsilon_0}$, and $k$ being the wave number in free space. $S_\mathrm{m}$ and $V_\mathrm{d}$ respectively represent the regions filled by conductors and the regions filled with dielectric materials in source region $V_0$, and they satisfy $\left(S_\mathrm{m}\bigcup V_\mathrm{d}\right)\subseteq V_0$. Compared with the formulations in [10], which only involves $\boldsymbol{J}_\mathrm{s}$, the last three terms in (5) are attributed to the introduction of equivalent volume current $\boldsymbol{J}_\mathrm{V}$. Considering (2), the second term after the second equal sign of

(3) can be rewritten as

$$\int_{V_0} \frac{1}{4}(\varepsilon_d - \varepsilon_0)|\boldsymbol{E}|^2 \, dV = \frac{1}{4\omega^2}\int_{V_d} \frac{|\boldsymbol{J}_V|^2}{(\varepsilon_d - \varepsilon_0)}\, dV. \quad (6)$$

Substituting (5) and (6) into (3), the DSFE in the source region can be evaluated by the current distributions directly.

Introducing the inner product of two vector functions $\boldsymbol{A}$ and $\boldsymbol{B}$ in $\Omega$, which is defined by $\langle \boldsymbol{A}, \boldsymbol{B} \rangle = \int_\Omega \boldsymbol{A} \cdot \boldsymbol{B}^* d\Omega$. Then the DSFE can be written as the sum of inner products

$$\Delta W = \langle \mathcal{L}^{\text{MM}}\boldsymbol{J}_S, \boldsymbol{J}_S \rangle + \langle \mathcal{L}^{\text{DM}}\boldsymbol{J}_V, \boldsymbol{J}_S \rangle$$
$$+ \langle \mathcal{L}^{\text{MD}}\boldsymbol{J}_S, \boldsymbol{J}_V \rangle + \langle \mathbb{Z}^{\text{DD}}\boldsymbol{J}_V, \boldsymbol{J}_V \rangle \quad (7)$$
$$= \begin{bmatrix} \boldsymbol{J}_S \\ \boldsymbol{J}_V \end{bmatrix}^H \begin{bmatrix} \mathcal{L}^{\text{MM}} & \mathcal{L}^{\text{DM}} \\ \mathcal{L}^{\text{MD}} & \mathbb{Z}^{\text{DD}} \end{bmatrix} \begin{bmatrix} \boldsymbol{J}_S \\ \boldsymbol{J}_V \end{bmatrix}.$$

Here the superscript H represents the Hermitian transpose of the matrix, and four integral operators are introduced

$$\mathcal{L}^{\text{MM}}\boldsymbol{J}_S = \frac{\eta}{16\pi c}\int_{S_m}\left\{ \boldsymbol{J}_S(\boldsymbol{r}') + \frac{1}{k^2}\left[\nabla' \cdot \boldsymbol{J}_S(\boldsymbol{r}')\right]\nabla \right\}\frac{\cos kR}{R}\, dS', \quad (8)$$

$$\mathcal{L}^{\text{DM}}\boldsymbol{J}_V = \frac{\eta}{16\pi c}\int_{V_d}\left\{ \boldsymbol{J}_V(\boldsymbol{r}') + \frac{1}{k^2}\left[\nabla' \cdot \boldsymbol{J}_V(\boldsymbol{r}')\right]\nabla \right\}\frac{\cos kR}{R}\, dV', \quad (9)$$

$$\mathcal{L}^{\text{MD}}\boldsymbol{J}_S = \frac{\eta}{16\pi c}\int_{S_m}\left\{ \boldsymbol{J}_S(\boldsymbol{r}') + \frac{1}{k^2}\left[\nabla' \cdot \boldsymbol{J}_S(\boldsymbol{r}')\right]\nabla \right\}\frac{\cos kR}{R}\, dS', \quad (10)$$

$$\mathbb{Z}^{\text{DD}}\boldsymbol{J}_V = \left(\mathcal{L}^{\text{DD}} + \boldsymbol{C}\right)\boldsymbol{J}_V, \quad (11)$$

where the constant vector operator $\boldsymbol{C}$ and the integral operator $\mathcal{L}^{\text{DD}}$ are defined by

$$\boldsymbol{C}\boldsymbol{J}_V = -\frac{1}{4\omega^2}\frac{\boldsymbol{J}_V(\boldsymbol{r}')}{(\varepsilon_d - \varepsilon_0)}, \quad (12)$$

$$\mathcal{L}^{\text{DD}}\boldsymbol{J}_V = \frac{\eta}{16\pi c}\int_{V_d}\left\{ \boldsymbol{J}_V(\boldsymbol{r}') + \frac{1}{k^2}\left[\nabla' \cdot \boldsymbol{J}_V(\boldsymbol{r}')\right]\nabla \right\}\frac{\cos kR}{R}\, dV'. \quad (13)$$

The source region $V_0$ is said to be resonant if (7) vanishes. In order to find the resonant frequency and the corresponding modal current distribution $\boldsymbol{J}$ (i.e., the surface current $\boldsymbol{J}_S$ and the equivalent volume current $\boldsymbol{J}_V$) that makes (7) vanish, a sufficient condition is

$$\begin{bmatrix} \mathcal{L}^{\text{MM}} & \mathcal{L}^{\text{DM}} \\ \mathcal{L}^{\text{MD}} & \mathbb{Z}^{\text{DD}} \end{bmatrix} \begin{bmatrix} \boldsymbol{J}_S \\ \boldsymbol{J}_V \end{bmatrix} = \boldsymbol{0}. \quad (14)$$

One can now follow the standard procedure of MoM [24] to solve (14). The currents may be expanded in terms of the RWG basis functions $\boldsymbol{f}_n^S$ [25] and SWG basis functions $\boldsymbol{f}_n^V$ [26]

$$\boldsymbol{J}_S(\boldsymbol{r}) = \sum_{n=1}^{N_S} I_n \boldsymbol{f}_n^S(\boldsymbol{r}), \qquad \boldsymbol{J}_V(\boldsymbol{r}) = \sum_{n=1}^{N_D} \Lambda_n \boldsymbol{f}_n^V(\boldsymbol{r}), \quad (15)$$

where $I_n$ and $N_S$ are the unknown coefficients of surface current and the total number of the unknowns in the conductor regions respectively; $\Lambda_n$ and $N_D$ are the unknown coefficients of the equivalent volume current and the total number of the unknowns in the dielectric region respectively. Introducing (15) into (14), we obtain the following real homogenous algebraic matrix equation by Galerkin method

$$\begin{bmatrix} l_{mn}^{\text{MM}} & l_{mn}^{\text{DM}} \\ l_{mn}^{\text{MD}} & Z_{mn}^{\text{DD}} \end{bmatrix} \begin{bmatrix} \boldsymbol{J}_S \\ \boldsymbol{J}_V \end{bmatrix} = \boldsymbol{0}, \quad (16)$$

where $[\boldsymbol{J}_S] = [I_1, I_2, \dots, I_{N_S}]^T$, $[\boldsymbol{J}_V] = [\Lambda_1, \Lambda_2, \dots, \Lambda_{N_D}]^T$ with the superscript T representing the transpose operation of the matrix, and the elements of the coefficient matrix are given by

$$l_{mn}^{\text{MM}} = \frac{\eta}{16\pi c}\int_{S_m}\int_{S_m}\boldsymbol{f}_m^S(\boldsymbol{r}) \cdot \boldsymbol{f}_n^S(\boldsymbol{r}')\frac{\cos kR}{R}\, dS'dS$$
$$- \frac{1}{k^2}\frac{\eta}{16\pi c}\int_{S_m}\int_{S_m}\nabla' \cdot \boldsymbol{f}_n^S(\boldsymbol{r}')\nabla \cdot \boldsymbol{f}_m^S(\boldsymbol{r})\frac{\cos kR}{R}\, dS'dS, \quad (17)$$

$$l_{mn}^{\text{DM}} = \frac{\eta}{16\pi c}\int_{S_m}\int_{V_d}\boldsymbol{f}_m^S(\boldsymbol{r}) \cdot \boldsymbol{f}_n^V(\boldsymbol{r}')\frac{\cos kR}{R}\, dV'dS$$
$$- \frac{1}{k^2}\frac{\eta}{16\pi c}\int_{S_m}\int_{V_d}\nabla \cdot \boldsymbol{f}_m^S(\boldsymbol{r})\nabla' \cdot \boldsymbol{f}_n^V(\boldsymbol{r}')\frac{\cos kR}{R}\, dV'dS, \quad (18)$$

$$l_{mn}^{\text{MD}} = \frac{\eta}{16\pi c}\int_{V_d}\int_{S_m}\boldsymbol{f}_m^V(\boldsymbol{r}) \cdot \boldsymbol{f}_n^S(\boldsymbol{r}')\frac{\cos kR}{R}\, dS'dV$$
$$- \frac{1}{k^2}\frac{\eta}{16\pi c}\int_{V_d}\int_{S_m}\nabla \cdot \boldsymbol{f}_m^V(\boldsymbol{r})\nabla' \cdot \boldsymbol{f}_n^S(\boldsymbol{r}')\frac{\cos kR}{R}\, dS'dV, \quad (19)$$

$$Z_{mn}^{\text{DD}} = -\frac{1}{4\omega^2}\int_{V_d}\frac{\boldsymbol{f}_m^V(\boldsymbol{r}) \cdot \boldsymbol{f}_n^V(\boldsymbol{r}')}{(\varepsilon_d - \varepsilon_0)}\, dV$$
$$+ \frac{\eta}{16\pi c}\int_{V_d}\int_{V_d}\boldsymbol{f}_m^V(\boldsymbol{r}) \cdot \boldsymbol{f}_n^V(\boldsymbol{r}')\frac{\cos kR}{R}\, dV'dV \quad (20)$$
$$- \frac{1}{k^2}\frac{\eta}{16\pi c}\int_{V_d}\int_{V_d}\nabla' \cdot \boldsymbol{f}_n^V(\boldsymbol{r}')\nabla \cdot \boldsymbol{f}_m^V(\boldsymbol{r})\frac{\cos kR}{R}\, dV'dV.$$

If the source region simply consists of conductors, the coefficient matrix in (16) will be reduced to a matrix containing the elements (17) only as discussed in [10]. Similarly, if the source region simply consists of dielectrics, the coefficient matrix in (16) will be reduced to a matrix containing the elements (20) only. The necessary and sufficient condition for the existence of a non-zero solution of (16) is that the determinant of its coefficient matrix is zero

$$\det[W] = \det\begin{bmatrix} l_{mn}^{\text{MM}} & l_{mn}^{\text{DM}} \\ l_{mn}^{\text{MD}} & Z_{mn}^{\text{DD}} \end{bmatrix} = 0. \quad (21)$$

The above equation determines the resonant frequencies $\omega$, and the corresponding resonant current modes can be found from (16). It is noted that (21) is a sufficient condition that makes (7) vanish. It can be shown that (21) is also a necessary condition and a proof is given in the Appendix.

Clearly, the results obtained from RMT only contain the resonant current modes that inherently exist in the source region, and no redundant non-resonant modes (such as inductive modes and capacitive modes in TCM) are involved. The patterns of these inherent resonant modes help determine which mode(s) to be excited and where to place the excitation sources. For these reasons, RMT provides a clear physical insight in resonant antenna design.

### B. Numerical Procedures of Extended RMT

To evaluate the resonant modes in a source region, the numerical procedure is implemented with an in-house MATLAB code and may be summarized into three steps.

1) *Search the approximate locations of the resonant frequencies*: The determinant (21) is calculated in a selected frequency range and the locations of the resonant frequencies in the range are determined approximately based on the sign changes of the determinant.

2) *Determine the precise resonant frequencies*: The bisection method [27] is applied to find the precise solutions of (21) on the basis of the approximated locations obtained in Step 1). In this paper, the precision of bisection method for all numerical examples are set as 0.01.

3) *Calculate the resonant modes and far-field patterns*: Once the resonant frequencies are determined in Step 2), the resonant modes are obtained from (16) by using the *null* function in MATLAB. The far-field patterns can be determined from the integral representation of the fields.

## III. NUMERICAL EXAMPLES

To verify the extended RMT, three numerical examples, including a cylindrical dielectric resonator antenna (DRA), a dual-band dielectric coated wire antenna, and a rectangular patch antenna, will be investigated. The numerical results will be compared with those obtained from other theories, the finite-difference-time-domain-solver in CST [28] and MoM-VSIE-solver in FEKO [29].

### A. Cylindrical dielectric resonator antenna

A dielectric cylinder that has been used as a numerical example in [31], [19], and [32] will be studied by the extended RMT. It will be assumed that the radius of the dielectric cylinder is 5.25 mm, the height is 4.6 mm, and the relative permittivity is $\varepsilon_r = 38$. The mesh size of the cylinder is set as $\lambda_d / 6$ ($\lambda_d$ represents the wavelength in the dielectric region), which results in 4227 tetrahedrons with 8851 unknowns.

For a dielectric cylinder, the resonant frequencies can be obtained from (21), which only includes the elements in (20). From (21), eight resonant modes are found from 4.5 GHz to 8.0 GHz and listed in the last row of TABLE I to compare with the numerical results obtained from other methods: the measurements in [30], the determinant root seeking method in a complex frequency domain [31], the method based on TCM-SIE in [19], and the method based on TCM-SIE and TCM-VIE in [32]. In TABLE I, the numbers listed inside the round brackets represent the degeneracy of the mode. It is observed that the resonant frequencies obtained by the extended RMT are very close to those from other methods. The errors of the extended RMT compared to the measurements are all less than 0.21%.

The distributions of electrical fields (**E**-fields) for the eight resonant modes and the corresponding radiation patterns are depicted in Fig. 2. The first and sixth modes are $TE_{01\delta}$ and $TM_{01\delta}$ modes, respectively. Other modes are classified as HEM modes. As compared with the modal fields and radiation patterns in [31] and [32], good agreements are observed.

TABLE I
THE RESONANT FREQUENCY AND DEGENERACY OBTAINED FROM DIFFERENT METHODS (IN GHz)

| Method | $TE_{01\delta}$ | $HEM_{11\delta}$ | $HEM_{12\delta}$ | $TM_{01\delta}$ | $HEM_{21\delta}$ |
|---|---|---|---|---|---|
| Measurement in [30] | 4.85 | NULL | 6.64 | 7.60 | 7.81 |
| Determinant Root Seeking in [31] | 4.83 | 6.33 | 6.63 | 7.52 | 7.75 |
| TCM-SIE in [19] | 4.88 | 6.30(2) | 6.68(2) | 7.50 | 7.75(2) |
| TCM-SIE in [32] | 4.87 | 6.35(2) | 6.64(2) | 7.53 | 7.72(2) |
| TCM-VIE in [32] | 4.90 | 6.45(2) | 6.75(2) | 7.80 | 8.00(2) |
| **Extended RMT** | **4.86** | **6.38(2)** | **6.65(2)** | **7.60** | **7.80(2)** |

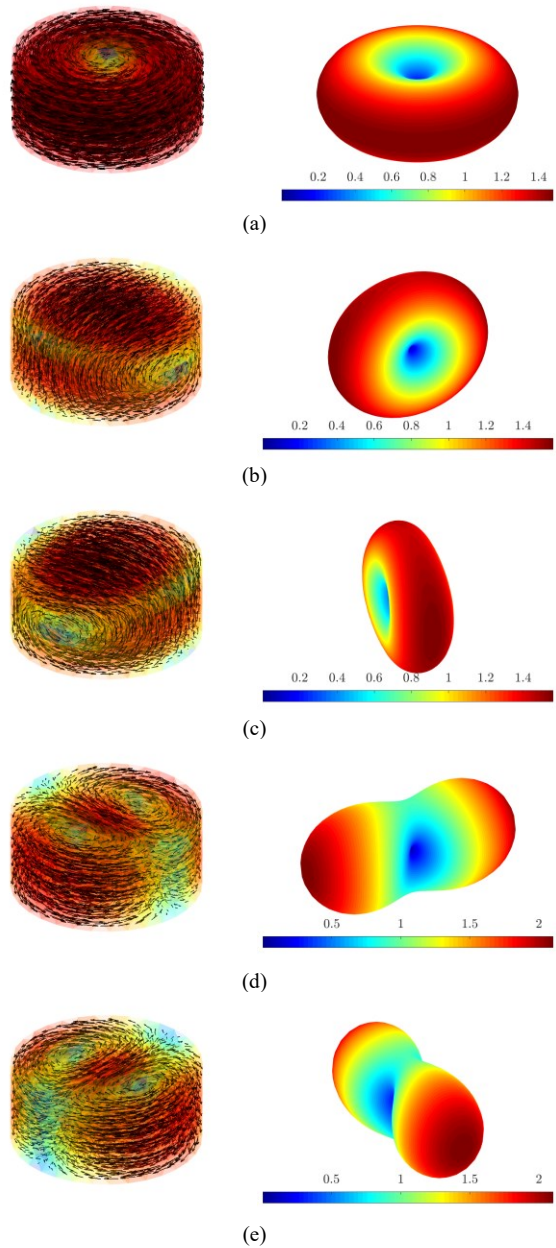

(a)

(b)

(c)

(d)

(e)

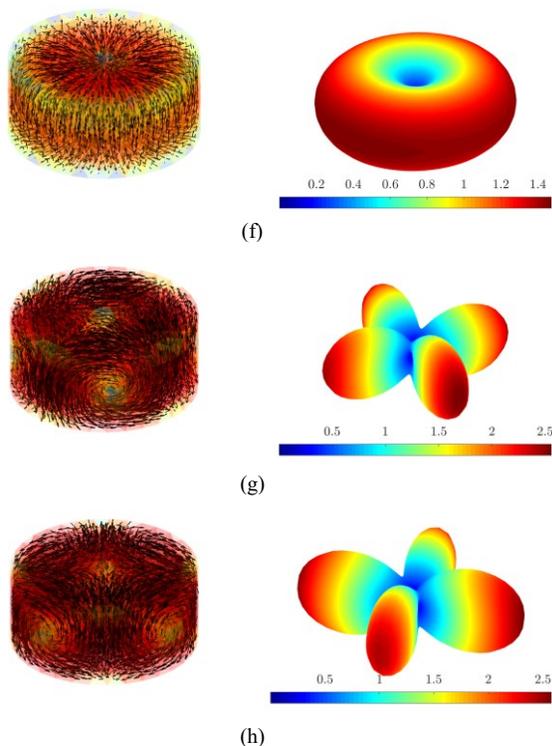

(f)

(g)

(h)

Fig. 2 Resonant $\boldsymbol{E}$-field distributions and the radiation patterns of the isolated cylindrical dielectric resonator. (a) $TE_{01\delta}$ at 4.86 GHz. (b) $HEM_{11\delta}$ at 6.38 GHz. (c) $HEM_{11\delta}$ at 6.38 GHz. (d) $HEM_{12\delta}$ at 6.65 GHz. (e) $HEM_{12\delta}$ at 6.65 GHz. (f) $TM_{01\delta}$ at 7.60 GHz. (g) $HEM_{21\delta}$ at 7.80 GHz. (h) $HEM_{21\delta}$ at 7.80 GHz.

To excite the $TE_{01\delta}$ mode in Fig. 2(a), one can introduce a conformal metal strip on the surface of the cylindrical dielectric resonator, as shown in Fig. 3. The length and width of the conformal metal strip (the region with orange color in Fig. 3) are 5 mm and 0.5 mm, respectively. The mesh size of the dielectric region remains unchanged, and the mesh size of the conformal metal strip is set as $\lambda_d / 12$. The source region consisting of the metal strip and the dielectric cylinder is discretized as 4182 tetrahedrons and 10 triangles. Due to the introduction of the conformal metal strip, the resonant modes of the source region will be different from the single dielectric cylinder and must therefore be recalculated. From (21), only one resonant mode is found from 4.5 GHz to 5.5 GHz, and the resonant frequency is 4.778 GHz. The corresponding resonant $\boldsymbol{E}$-field distribution and radiation pattern are the same with those in Fig. 2(a), and are shown in Fig. 4.

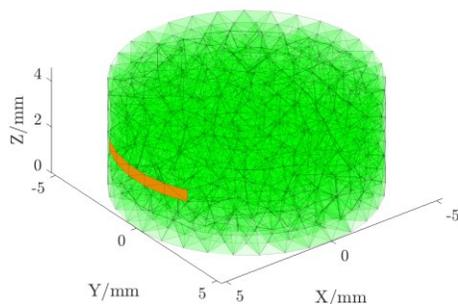

Fig. 3 Composite structure of the cylindrical dielectric resonator combined with a conformal metallic strip.

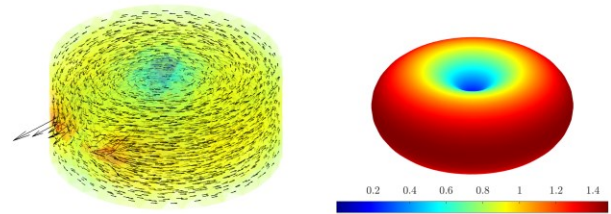

Fig. 4 Resonant $\boldsymbol{E}$-field distribution and the radiation pattern of above metallic-dielectric composite structure calculated by the extended RMT formulations.

To further verify the results obtained from the extended RMT, a feeding point is introduced in the middle of the metal strip and the cylindrical DRA is simulated by a lumped port placed at the feeding gap in FEKO and CST. It can be seen from the reflection coefficients in Fig. 5 that the cylindrical DRA is resonant at 4.80 and 4.783 GHz in FEKO and CST, respectively, while the resonant frequency obtained from the extended RMT is 4.778 GHz. The $\boldsymbol{E}$-field distributions and the radiation patterns of the cylindrical DRA obtained from FEKO and CST are displayed in Fig. 6. Comparing Fig. 6 with Fig. 4, it is clear that the $\boldsymbol{E}$-field distributions and radiation patterns obtained from the extended RMT formulations agree well with those from the full-wave simulations with practical excitations.

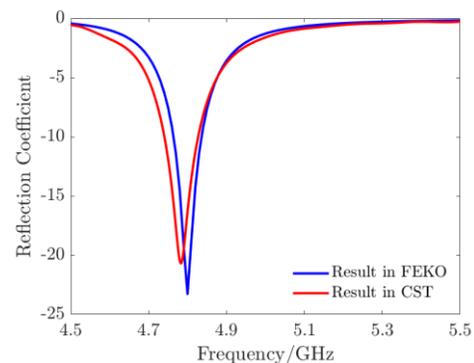

Fig. 5 Comparison of the simulated reflection coefficients of the cylindrical DRA in FEKO and CST.

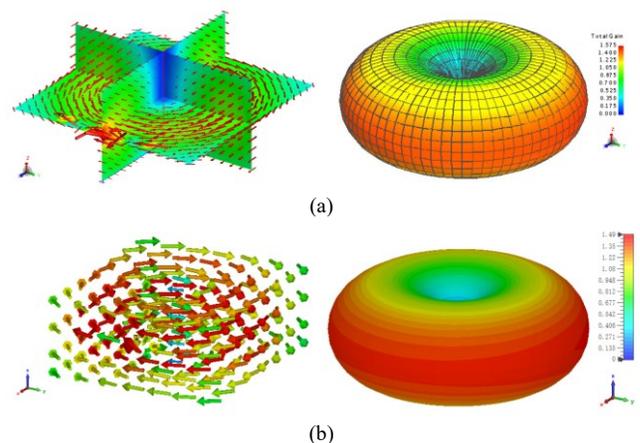

(a)

(b)

Fig. 6 $\boldsymbol{E}$-field distributions and the radiation patterns obtained from (a) FEKO. (b) CST.

### B. Dual-band dielectric-Coated Wire Antenna

The second example is a dielectric-coated metallic wire shown in Fig. 7. The radius of the metallic wire is $r_1 = 3.175$ mm, and the length is 250 mm. The metallic wire is coated with a dielectric material of relative permittivity $\varepsilon_r = 3.2$ whose outer radius is set as $r_2 = 6.35$ mm. The dielectric coated metallic wire is discretized as 426 triangles in metallic region (the meshes in orange color in Fig. 7) and 1449 tetrahedrons in dielectric region (the meshes in green color in Fig. 7).

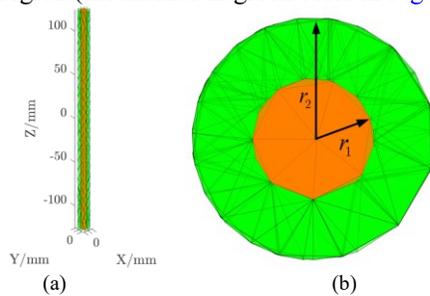

Fig. 7 The dielectric-coated metallic wire. (a) Meshes. (b) Top view of the meshed structure.

From (21), the first three resonant frequencies are found to be 0.492 GHz, 1.012 GHz, and 1.531 GHz. The corresponding resonant current distributions and the radiation patterns of the dielectric-coated metallic wire are plotted in Fig. 8. As compared with the resonant frequencies and the modal fields obtained from TCM in [33], good agreements are observed. In Fig. 8 (a), the first modal current distribution at 0.492 GHz is along the metallic wire and has strongest current at the center of the wire, and thus can be considered as a half-wavelength dipole. The second modal current distribution in Fig. 8 (b) corresponds to a full-wavelength dipole, while the third modal current distribution in Fig. 8 (c) corresponds to a one-and-half -wavelength dipole. Therefore, the first and third resonant modes can be effectively excited if a voltage source is placed at the center of the wire.

Through exciting the first and third resonant modes calculated from the extended RMT, a dual-band dielectric-coated wire antenna can be realized with a center-driven excitation. The simulated reflection coefficients of the dual-band dielectric-coated wire antenna with FEKO and CST is shown in Fig. 9, in which two reflection zeros occur at 0.50 and 1.55 GHz in FEKO and two reflection zeros occur at 0.50 and 1.51 GHz in CST, while the extended RMT shows the two resonant frequencies at 0.492 GHz and 1.531 GHz. Fig. 10 and Fig. 11 demonstrate the resonant current distributions and radiation patterns obtained from different full-wave simulations with a excitation, which are all consistent with the results shown in Fig. 8. Therefore, the extended RMT is able to provide a clearer physical insight into the working mechanism of the dual-band dielectric-coated wire antenna.

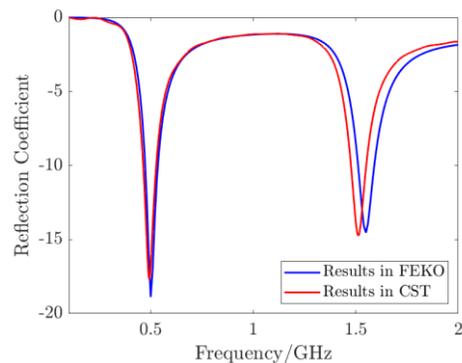

Fig. 9 Comparison of the simulated reflection coefficients of the dual-band dielectric-coated wire antenna with FEKO and CST.

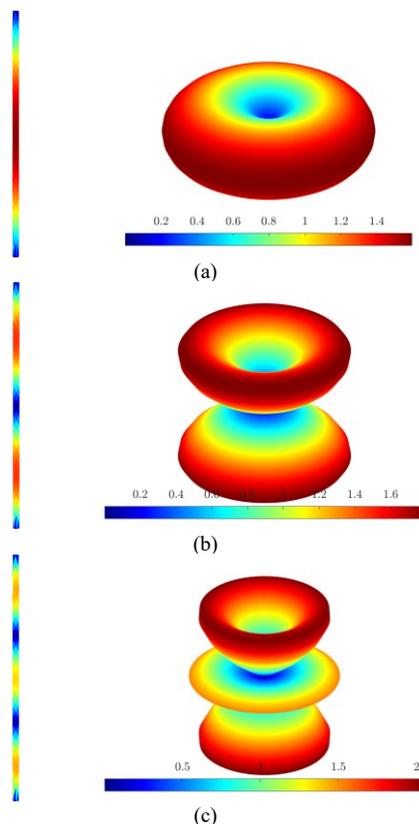

Fig. 8 Resonant current distributions and the radiation patterns of the dielectric-coated metallic wire at (a) 0.492 GHz. (b) 1.012 GHz. (c) 1.531 GHz.

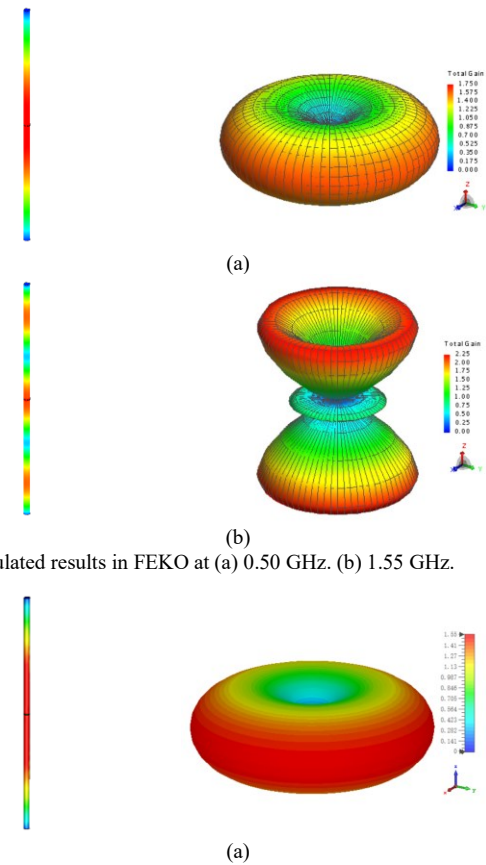

Fig. 10 Simulated results in FEKO at (a) 0.50 GHz. (b) 1.55 GHz.

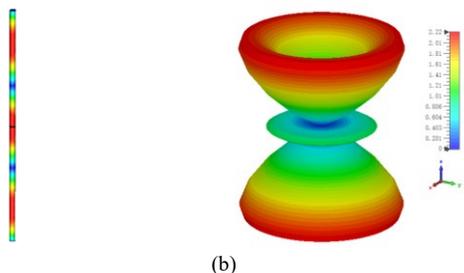

Fig. 11 Simulated results in CST at (a) 0.50 GHz. (b) 1.51 GHz.

## C. Rectangular patch antenna

We now investigate the modal behaviors of the rectangular patch antenna. The first four resonant modes are calculated by using the cavity model in [34]. The width and length of the patch are 50 mm and 76 mm, respectively. The patch is mounted on a grounded dielectric substrate with a relative permittivity of $\varepsilon_r = 3.38$ and a height of 1.524 mm. The bottom surface of the substrate is the ground plane, whose width and length are 100 mm and 152 mm, respectively. The source region, consisting of the patch, the substrate and the ground, is discretized as 1728 triangles and 3400 tetrahedrons.

From (21), eight resonant modes are found to be 0.861 GHz, 1.088 GHz, 1.522 GHz, 1.607 GHz, 1.878 GHz, 1.978 GHz, 2.010 GHz, and 2.192 GHz, respectively. The corresponding modal current distributions are depicted in Fig. 12. The resonant current distributions in Fig. 12 (b), Fig. 12 (d), Fig. 12 (f), and Fig. 12 (h) are recognized as $TM_{10}$, $TM_{01}$, $TM_{11}$, and $TM_{20}$ modes as discussed in [34], and they are in good agreement with the results obtained from TCM based on the mixed potential integral equation (MPIE) [35]. The resonant frequencies of those four modes are listed in the last column of TABLE II to compare with the numerical results obtained from the cavity model [34], MPIE-based TCM [35], and the sub-structure TCM method that relies on VSIE [1]. One can see that the resonant frequencies obtained from the extended RMT also agree well with the results from other methods. Note that these four resonant modes can be easily excited [20].

The current distributions of the first and third resonant modes presented in Fig. 12(a) and Fig. 12(c) mainly concentrate on the ground plane and the currents on the patch are negligible. Fig. 12(e) and Fig. 12 (g) indicate that the intensities of the fifth and seventh resonant current distributions on the ground plane and on the patch are comparable, and they are generated by the combination of the patch and ground plane. The designations and applications of these four modes remain to be explored.

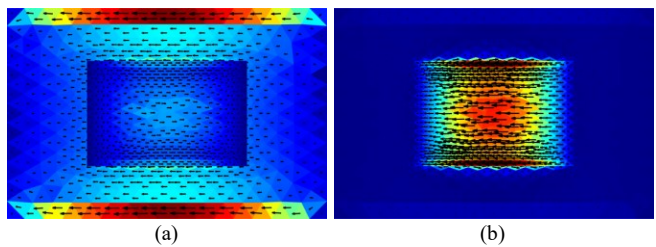

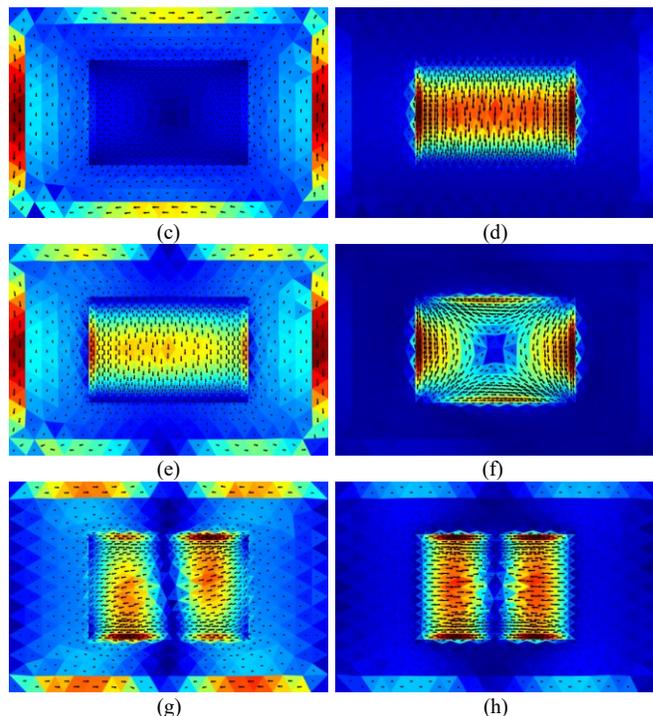

Fig. 12 The current distributions of the rectangular patch antenna at resonant frequencies. (a) 0.861 GHz. (b) 1.088 GHz. (c) 1.522 GHz. (d) 1.607 GHz. (e) 1.878 GHz. (f) 1.978 GHz. (g) 2.010 GHz. (h) 2.192 GHz.

TABLE II
RESONANCE FREQUENCIES (IN GHZ) OF THE RECTANGULAR MICROSTRIP
PATCH ANTENNA OBTAINED FROM DIFFERENT METHODS

| Mode | Cavity model | MPIE based TCM | Sub-structure TCM based VSIE | **Extended RMT** |
|------|------|------|------|------|
| $TM_{10}$ | 1.075 | 1.08 | 1.09 | **1.088** |
| $TM_{01}$ | 1.605 | 1.60 | 1.65 | **1.607** |
| $TM_{11}$ | 1.955 | 1.95 | 2.03 | **1.978** |
| $TM_{20}$ | 2.145 | 2.13 | 2.19 | **2.192** |

## IV. CONCLUSION

In this paper, the RMT is extended to an arbitrary source region, consisting of conductors, dielectrics, or a combination of both. In contrast to the modal decomposition in TCM, in which the redundant non-resonant modes (inductive modes and capacitive modes) are always naturally included, the RMT only addresses the resonant modes that are most relevant to antenna design. Once the resonant modes are determined by the RMT, the rest is to use various excitations to realize the resonant current modes in the selected source region and frequency band. Distinguished from the other modal theories started from the field relationship between the incident and scattering field, the formulations of RMT are directly derived from the definition of resonance in terms of the DSFE, and all the resonant frequencies of a source region can be obtained by finding the roots of a determinant in real frequency domain. Numerical results demonstrate that the extended RMT not only provides accurate solutions for resonant modes but also offers a clearer physical picture about the working principles of resonant antennas to better understand the intrinsic resonances of complicated antenna structures, which is of vital importance in antenna designs.

Although the materials investigated in this work only involve metals and dielectrics, other materials, such as the magnetic materials, can also be included by introducing equivalent magnetic currents in the formulation of RMT through the use of compensation theorem [3], [23]. In this respect, much remains to be explored.

## Appendix

In this Appendix, we provide a proof that (21) is also a necessary condition that makes (7) vanish.

Since $\varepsilon_d - \varepsilon_0$ in (12) is a scalar, the operator $\boldsymbol{C}$ is a symmetric operator. It has been demonstrated in [10] that the operator $\mathcal{L}^{\mathrm{MM}}$ is symmetric, and the operator $\mathcal{L}^{\mathrm{DD}}$, which has a form similar to $\mathcal{L}^{\mathrm{MM}}$, is also symmetric. Therefore, the operator $\mathbb{Z}^{\mathrm{DD}}$ is also symmetric.

After discretization, (7) can be written as a quadratic form

$$\begin{bmatrix} \boldsymbol{J}_{\mathrm{S}} \\ \boldsymbol{J}_{\mathrm{V}} \end{bmatrix}^{\mathrm{H}} \begin{bmatrix} \mathcal{L}^{\mathrm{MM}} & \mathcal{L}^{\mathrm{DM}} \\ \mathcal{L}^{\mathrm{MD}} & \mathbb{Z}^{\mathrm{DD}} \end{bmatrix} \begin{bmatrix} \boldsymbol{J}_{\mathrm{S}} \\ \boldsymbol{J}_{\mathrm{V}} \end{bmatrix} = [\boldsymbol{J}]^{\mathrm{H}} [W] [\boldsymbol{J}], \qquad (22)$$

where $[\boldsymbol{J}] = [\boldsymbol{J}_{\mathrm{S}}, \boldsymbol{J}_{\mathrm{V}}]^{\mathrm{T}}$. The reciprocity theorem [22] gives

$$\int_{S_m} \boldsymbol{J}_{\mathrm{S}} \cdot \boldsymbol{E}_{\mathrm{V}} \mathrm{d}S = \int_{V_d} \boldsymbol{J}_{\mathrm{V}} \cdot \boldsymbol{E}_{\mathrm{S}} \mathrm{d}V, \qquad (23)$$

where $\boldsymbol{E}_{\mathrm{S}}$ and $\boldsymbol{E}_{\mathrm{V}}$ represent the fields produced by the surface current $\boldsymbol{J}_{\mathrm{S}}$ and the equivalent volume source $\boldsymbol{J}_{\mathrm{V}}$ respectively. Since the current distribution $\boldsymbol{J}_{\mathrm{S}}$ and $\boldsymbol{J}_{\mathrm{V}}$ considered in this paper are real, it follows that

$$\int_{S_m} \boldsymbol{J}_{\mathrm{S}}^* \cdot \boldsymbol{E}_{\mathrm{V}} \mathrm{d}S = \int_{V_d} \boldsymbol{J}_{\mathrm{V}}^* \cdot \boldsymbol{E}_{\mathrm{S}} \mathrm{d}V. \qquad (24)$$

Similarly,

$$\int_{S_m} \boldsymbol{J}_{\mathrm{S}}^* \cdot \mathrm{Im}(\boldsymbol{E}_{\mathrm{V}}) \mathrm{d}S = \int_{V_d} \boldsymbol{J}_{\mathrm{V}}^* \cdot \mathrm{Im}(\boldsymbol{E}_{\mathrm{S}}) \mathrm{d}V, \qquad (25)$$

which can be rewritten as $\left\langle \mathcal{L}^{\mathrm{DM}} \boldsymbol{J}_{\mathrm{V}}, \boldsymbol{J}_{\mathrm{S}} \right\rangle = \left\langle \mathcal{L}^{\mathrm{MD}} \boldsymbol{J}_{\mathrm{S}}, \boldsymbol{J}_{\mathrm{V}} \right\rangle$. Therefore, the matrix $[W]$ in (21) is symmetric, and there exists an orthogonal matrix $[O]$ such that

$$[O]^{-1} [W] [O] = [D], \qquad (26)$$

where $[D] = [\lambda_1, \lambda_2, \cdots, \lambda_{N_S + N_D}]^{\mathrm{T}}$ is a diagonal matrix. Introducing a new vector $[Y] = [y_1, y_2, \cdots, y_{N_S + N_D}]^{\mathrm{T}}$ defined by

$$[\boldsymbol{J}] = [O][Y],$$

and substituting this into (22), we obtain

$$\begin{bmatrix} \boldsymbol{J}_{\mathrm{S}} \\ \boldsymbol{J}_{\mathrm{V}} \end{bmatrix}^{\mathrm{H}} \begin{bmatrix} \mathcal{L}^{\mathrm{MM}} & \mathcal{L}^{\mathrm{DM}} \\ \mathcal{L}^{\mathrm{MD}} & \mathbb{Z}^{\mathrm{DD}} \end{bmatrix} \begin{bmatrix} \boldsymbol{J}_{\mathrm{S}} \\ \boldsymbol{J}_{\mathrm{V}} \end{bmatrix} = [Y]^{\mathrm{T}} [O]^{\mathrm{T}} [O] [D] [Y]$$

$$= [Y]^{\mathrm{T}} [D] [Y] = \sum_{n=1}^{N} \lambda_n y_n^2, \qquad (27)$$

where $[O]^{\mathrm{T}} [O] = [I]$, $[I]$ is the unit matrix of order $N_S + N_D$, and $N = N_S + N_D$. If the above quadratic form is required to be zero for an arbitrary current distribution $\boldsymbol{J}$, we must have $\lambda_n = 0 \, (n = 1, 2, \cdots, N_S + N_D)$. This implies

$$\det[W] = \det[D] = \prod_{n=1}^{N} \lambda_n = 0. \qquad (28)$$

Thus we have proved that (21) is also a necessary condition that make (7) vanish.

## References

[1] R. B. Adler, L. J. Chu, R. M. Fano, *Electromagnetic Energy Transmission and Radiation*, New York, John Wiley & Sons,1960.

[2] R. E. Collin, *Foundations for Microwave Engineering*, New York: McGraw Hill, 1966.

[3] W. Geyi, *Foundations of Antenna Radiation Theory: Eigenmode Analysis*. John Wiley & Sons, 2023.

[4] C. E. Baum, "On the singularity expansion method for the solution of electromagnetic interaction problems," *AFWL, Interaction Notes*, vol.88, Dec. 1971.

[5] C. E. Baum, "On the eigenmode expansion method for electromagnetic scattering and antenna problems, part I: Some basic relations for eigenmode expansions and their relation to the singularity expansion," *Interaction Note*, vol. 229, Jan. 13, 1975, Air Force Weapons Lab., Kirtland Air Force Base, Albuquerque, NM, USA.

[6] R. J. Garbacz, "Modal expansions for resonance scattering phenomena," *Proc. IEEE*, vol. 53, no. 8, pp. 856–864, Aug. 1965.

[7] R. J. Garbacz and R. Turpin, "A generalized expansion for radiated and scattered fields," *IEEE Trans. Antennas Propag.*, vol. 19, no. 3, pp. 348–358, May 1971.

[8] R. F. Harrington and J. R. Mautz, "Theory of characteristic modes for conducting bodies," *IEEE Trans. Antennas Propag.*, vol. 19, no. 5, pp. 622–628, Sep. 1971.

[9] R. F. Harrington and J. R. Mautz, "Computation of characteristic modes for conducting bodies," *IEEE Trans. Antennas Propag.*, vol. 19, no. 5, pp. 629–639, Sep. 1971.

[10] R. Xiao, G. Wen, and W. Wu, "Theory of Resonant Modes and its Application," *IEEE Access*, vol. 9, pp. 114945-114956, 2021.

[11] W. Sun et al, "Determination of the natural modes for a rectangular plate," *IEEE Trans. Antennas Propag.*, vol. 38, no.5, pp. 643-652, May 1990.

[12] Y. Long, "Determination of the natural frequencies for conducting rectangular boxes," *IEEE Trans. Antennas Propag.*, vol. AP-42, no. 7, pp. 1016–1021, 1994.

[13] R. F. Harrington, J. R. Mautz, Y. Chang, "Characteristic modes for dielectric and magnetic bodies," *IEEE Trans. Antennas Propag.*, vol. 20, no. 2, pp. 194–198, Mar. 1972.

[14] Y. Chang, R. F. Harrington, "A surface formulation for characteristic modes of material bodies," *IEEE Trans. Antennas Propag.*, vol. 25 no. 6, pp. 789–795, Nov. 1977.

[15] P. Ylä-Oijala, "Generalized theory of characteristic modes," *IEEE Trans. Antennas Propag.*, vol. 67, no. 6, pp. 3915-3923, Jun. 2019.

[16] Q. Wu, "General metallic-dielectric structures: A characteristic mode analysis using volume-surface formulations," *IEEE Antennas Propag. Mag.*, vol. 61, no. 3, pp. 27-36, Jun. 2019.

[17] M. Gustafsson, L. Jelinek, K. Schab, and M. Capek, "Unified theory of characteristic modes: Part I – Fundamentals," *IEEE Trans. Antennas Propag.*, vol. 70, pp. 11801–11813, December 2022.

[18] M. Capek, J. Lundgren, M. Gustafsson, K. Schab, and L. Jelinek, "Characteristic mode decomposition using the scattering dyadic in arbitrary full-wave solvers," *IEEE Trans. Antennas Propag.*, vol. 71, pp. 830–839, January 2023.

[19] Y. Chen, "Alternative surface integral equation-based characteristic mode analysis of dielectric resonator antennas," *IET Microw., Antennas Propag.*, vol. 10, no. 2, pp. 193–201, Jan. 2016.

[20] Y. Chen, and C.-F. Wang, *Characteristic Modes: Theory and Applications* in Antenna Engineering. Hoboken, NJ, USA: Wiley, 2015.

[21] M. Capek and K. Schab, "Computational aspects of characteristic mode decomposition: An overview," *IEEE Antennas Propag. Mag.*, vol. 64, no. 2, pp. 23–31, Apr. 2022.

[22] R. F. Harrington, *Time-Harmonic Electromagnetic Fields*. Hoboken, NJ, USA: Wiley, 2001.

[23] B. D. Popvic, "Electromagnetic field theorems," *IEE Proc. On Science Measurement and Technology*, Vol. 128, Pt. A, 47–63, Jan. 1981.

[24] R. F. Harrington, *Field Computation by Moment Methods*. Hoboken, NJ, USA: Wiley, 1993.

[25] S. M. Rao, D. R. Wilton, and A. W. Glisson, "Electromagnetic scattering by surfaces of arbitrary shape," *IEEE Trans. Antennas Propag.*, vol. 30, no. 3, pp. 409-418, May. 1982.

[26] D. H. Schaubert, D. R. Wilton, and A. W. Glisson, "A tetrahedral modeling method for electromagnetic scattering by arbitrarily shaped inhomogeneous dielectric bodies," *IEEE Trans. Antennas and Propagation*, vol. 32, no. 1, pp. 77-85, January 1984.

[27] W. H. Press, S. A. Teukolsky, W. T. Vetterling, and B. P. Flannery, *Numerical Recipes in C*. Cambridge, U.K.: Cambridge Univ. Press, 1992.

[28] CST Computer Simulation Technology AG., USA, (Oct. 2018). [Online]. Available: https://www.cst.com.

[29] FEKO. Altair. (2019) [Online]. Available: https://altairhyperworks. com/product/Feko.

[30] A. W. Glisson, D. Kajfez, and J. James, "Evaluation of modes in dielectric resonators using a surface integral equation formulation," *IEEE Trans. Microw. Theory Techn.*, vol. 31, no. 12, pp. 1023–1029, Dec. 1983.

[31] D. Kajfez, A. W. Glisson, and J. James, "Computed modal field distributions for isolated dielectric resonators," *IEEE Trans. Microw. Theory Techn.*, vol. 32, no. 12, pp. 1609–1616, Dec. 1984.

[32] R. Lian, J. Pan, and S. Huang, "Alternative surface integral equation formulations for characteristic modes of dielectric and magnetic bodies," *IEEE Trans. Antennas Propag.*, vol. 65, no. 9, pp. 4706–4716, Sept. 2017.

[33] S. D. Huang, C. F. Wang, J. Pan, D. Q. Yang, and M. C. Tang, "Full equiphase characteristic mode solution to lossless composite metallic dielectric problems," *IEEE Trans. Antennas Propag.*, vol. 69, no. 12, pp. 8526–8538, Dec. 2021.

[34] Z. N. Chen and M. Y. W. Chia, *Broadband Planar Antennas*: *Design and Applications*. Hoboken, NJ, USA: Wiley, 2006.

[35] Y. Chen, L. Guo, and S. Yang, "Mixed-potential integral equation based characteristic mode analysis of microstrip antennas," *Int. J. of Antennas Propag.*, vol. 2016, pp. 1–8, 2016.

[36] S. Huang, M.-C. Tang, and C.-F. Wang, "Accurate modal analysis of microstrip patch antennas using sub-structure characteristic modes," in *2021 International Applied Computational Electromagnetics Society (ACES-China) Symposium*. IEEE, 2021, pp. 1-2.